\documentclass[aps,amssymb,showkeys]{revtex4}
\usepackage{amsmath}
\usepackage{graphicx}

\usepackage[pdftex, bookmarks, colorlinks=true, plainpages = false, citecolor = red, urlcolor = blue, filecolor = blue]{hyperref}

\newcommand{\be}{\begin{equation}}
\newcommand{\ee}{\end{equation}}
\newcommand{\bea}{\begin{eqnarray}}
\newcommand{\eea}{\end{eqnarray}}
   \def\l{\lambda}   \def\G{\Gamma} 
\def\p{\partial}   \def\x{\xi}

\newcommand{\cF}{{\cal F}}
\newcommand{\gf}[1]{\Gamma\left(\frac{#1}{3}\right)}
\newcommand{\ph}[1]{\left(\frac{#1}{3}\right)_q}
\newcommand{\gfM}[1]{\Gamma\left(\frac{#1}{M+1}\right)}
\newcommand{\gft}[1]{\Gamma\left(\frac{#1}{2(M+1)}\right)}

\begin{document}

\title{First Column Boundary Operator Product Expansion Coefficients}
\date{\today}

\author{Jacob J. H. Simmons}
\email{j.simmons1@physics.ox.ac.uk} 
\affiliation{LASST and Department of Physics \& Astronomy,
University of Maine, Orono, ME 04469, USA}
\affiliation{Rudolf Peierls Centre for Theoretical Physics
1 Keble Road, Oxford OX1 3NP, UK}
\author{Peter Kleban}
\email{kleban@maine.edu}
\affiliation{LASST and Department of Physics \& Astronomy,
University of Maine, Orono, ME 04469, USA}

\begin{abstract} 
We calculate  boundary operator product expansion coefficients $C^{abc}_{mnr}$ for boundary operators $\phi^{abc}_{1,s}$ in 
the first column of the Kac table in conformal field theories, where $\{abc\}$ specify the boundary conditions. For  $c=0$ we give closed form expressions for all such coefficients. Then we generalize to the augmented minimal models, giving explicit expressions for  coefficients valid when $\phi_{1,2}$ mediates a change from fixed to free (or any two specified) boundary conditions.  These quantities are determined by computing an arbitrary  four-point correlation 
function $\langle \phi^{ab}_{1,2}\phi^{bc}_{1,s_1}\phi^{cd}_{1,s_2}\phi^{da}_{1,s_3} \rangle$.  Our calculation first determines the 
appropriate (non-logarithmic) conformal blocks by using standard null-vector methods. The behavior of 
these blocks under crossing symmetry then provides  a general closed form expression for the $C^{abc}_{mnr}$, as 
a product of ratios of $\G$ functions.  This calculation was inspired by the need for several of these 
coefficients in certain correlation function formulas for  critical two-dimensional percolation and the augmented
$q=2$ and $q=3$ state critical Potts models.
\end{abstract}
\maketitle

%SECTION
\section{Introduction}\label{intro}

Conformal field theory (CFT) \cite{BPZ,JCedge} allows the determination of several results for 2D critical 
models in regions with boundaries.   In the case of critical percolation these include crossing probabilities,   such 
as  Cardy's formula for the horizontal crossing probability \cite{JC1,SS}, Watt's formula for the 
horizontal-vertical crossing probability \cite{W,D}, the expected number of horizontal crossing clusters  
\cite{JC2,SS2}, and some new formulas for related crossing and other quantities \cite{KSZ,SKZ,SKZip}.  
Their derivations require the calculation of correlation functions of boundary condition changing (bcc) 
operators $\phi_{1,s}$ from the first column of the Kac table for the $c=0$ boundary CFT \cite{JC1}.  The 
resulting formulas include certain  bcc operator product expansion (OPE) coefficients $C_{mnr}$, which (among other things) we 
calculate here.  We first determine {\it all} $C_{mnr}$ of first-column Kac table operators $\phi_{1,s}$ 
at $c=0$ (here there is no dependence on boundary conditions).  As a byproduct, we find the (non-logarithmic) conformal blocks for an arbitrary four-point 
function $\langle \phi_{1,2}\phi_{1,s_1}\phi_{1,s_2}\phi_{1,s_3} \rangle$.  Then we generalize these 
results to the (augmented) minimal models, giving explicit  expressions for the $C^{abc}_{mnr}$ valid when $\phi_{1,2}$ mediates a change from fixed to free boundary conditions.  These are useful in formulas 
generalizing the exact factorizations for percolation found in \cite{KSZ,SKZip} to the Fortuin-Kasteleyn representation of the $q=2$ and $q=3$ 
state Potts models \cite{SSKZ}.  

%SECTION
\section{Boundary Operator Product Expansions Coefficients at $c=0$} \label{c=0}

In this section we determine certain conformal blocks and operator product expansion coefficients at 
$c=0$.  These results are generalized to the augmented minimal models in section \ref{min}.

In regions with a boundary there are three distinct classes of OPEs, namely bulk-bulk, boundary-boundary 
or bulk-boundary \cite{Lew}.  In this article we consider  the boundary-boundary case exclusively.   
Correlation functions containing bcc operators that belong to the Kac table obey differential equations, 
which allow us to determine them in closed form.  From their solutions we can deduce the correlation 
functions and (our purpose here) the boundary OPE coefficients.   As mentioned, our motivation is that 
these quantities arise in problems at the percolation point. 

We begin with the $\phi_{1,2}$ Kac operator, which obeys the fusion rule
\be \label{fus}
[\phi_{1,2}] \times [\phi_{1,s}] = [\phi_{1,s-1}] +[\phi_{1,s+1}]\; .
\ee
Repeatedly applying (\ref{fus}) generates all Kac operators  $\phi_{1,s}$.  This set of operators is 
closed under OPEs and constitutes the only elements of the Kac table that we consider.  Note that 
$\phi_{1,2}$ generates a ``fixed to free" bcc in critical percolation \cite{JC1}, which is the origin of 
its interest for us.

Since the $\phi_{1,s}$ only depend on one parameter, we introduce the notation
\be
\psi_m := \phi_{1,m+1}\; ,
\ee
with $m \in {\mathbf N}$.
The index $m$ specifies the number of $\psi_1$ fusions  required to generate $\psi_m$. This includes the 
identity ($\psi_0$).  In applications to percolation, the free-to-fixed operator is $\psi_1$, while 
$\psi_2$ generates a cluster anchored at a point \cite{KSZ} and a physical interpretation of $\psi_3$ is 
given in  \cite{SKZ}.
From the Kac weight formula we find that (at $c=0$) the weights are given by 
\be
h_m=\frac{m(m-1)}{6}=\frac{1}{3}\binom{m}{2}\; .
\ee
Note that the integral values of $h_m$ are the Euler pentagonal numbers.

The form of the OPE that we use in this paper (here including boundary condition dependence) is
\be\label{genOPE}
\psi^{ab}_i(\l) \psi^{bc}_j(0)=\sum_{\substack{p\, = |i-j|\\ p- |i-j| \: \mathrm{even}}}^{i+j}C^{abc}_{i j 
p}\l^{h_p-h_i-h_j}\psi^{ac}_p(0) + \mathcal{O}(\l)^{1+h_p-h_i-h_j}\; .
\ee
The sum includes all the primary operators dictated by the fusion rules for the $\psi_m$, and 
$\mathcal{O}(z)$ includes all contributions from the descendants of the primary fields.  

In a general theory the boundary OPEs  depend on the boundary conditions assigned to each interval between 
the bcc operators. Thus there are additional indices on the OPE coefficients to represent the boundary 
conditions (see  \cite{Lew}).  For percolation there are only two types of conformally invariant boundary 
conditions, free or fixed (see \cite{JC1, SKZ}).  However, they are equivalent by duality.  As a 
result, the magnitudes of the OPE coefficients are 
independent of boundary conditions and their indices on the $C_{mnr}$ are not needed.  In addition, the normalizations of boundary operator two-point functions are independant of boundary conditions and we assume, with out loss of generality, that these two point functions are all normalized with coefficient unity, 
$$
\langle \psi_m(x_2) \psi_m(x_1) \rangle = (x_2-x_1)^{-2 h_m}\; ,
$$
(see sections \ref{anal} and \ref{min} for further details on this point).

To extract the OPE coefficients, we make use of four-point correlation functions  
\be \label{CorrFunc}
\lim_{\x \to \infty} \x^{2 h_l}\langle \psi_i(0) \psi_j(\l) \psi_k(1) \psi_l(\x) \rangle=\sum_p C_{i j p} 
C_{p k l}\mathcal{F}_{i j, k l}^p(\l) \;.
\ee
The functions $\cF_{i j, k l}^p(\l)$ are called conformal blocks.  They are the contributions to the 
correlation function from individual conformal families appearing in the OPE (\ref{genOPE}), and are 
normalized so that the coefficient of the leading term (as $\l \to 0$) is $1$.  The form of the 
correlation function in (\ref{CorrFunc})  follows on taking the limit $\l \to 0$ and using (\ref{genOPE}). 
 Alternatively, one may take the limit $\l \to 1$ and use the $\psi_k(1) \psi_j(\l)$ OPE. The equivalence 
of these two expressions yields the crossing symmetry relation \cite{BPZ}
\be\label{GenXSym}
\sum_p C_{i j p}C_{p k l}\mathcal{F}_{i j, k l}^p(\l)=\sum_q C_{k j q}C_{q i l} \mathcal{F}_{k j, i 
l}^q(1-\l) \;,
\ee
which we employ to determine the OPE coefficients. Note that the grouping of the indices in the conformal 
blocks is significant; however the indices of an OPE coefficient may be permuted without changing its 
value \cite{BPZ}. 

%SECTION
\section{Analysis} \label{anal}

We begin our analysis by calculating an arbitrary four-point function that includes at least one $\psi_1$,
\be \label{CorrFunc2}
\lim_{\x \to \infty} \x^{2 h_r}\langle \psi_m(0) \psi_1(\l) \psi_n(1) \psi_r(\x) \rangle = \sum_p C_{m 1 
p} C_{p n r}\mathcal{F}_{m 1, n r}^p(\l)\; .
\ee
Conformal covariance fixes the form of the four-point function up to a function of the cross-ratio 
\cite{BPZ}.  We choose the form
\bea \label{cfform}
\mathcal{G}_{m 1,n r}(x_1,x_2,x_3,x_4)&=&\langle \psi_m(x­_1) \psi_1(x_2) \psi_n(x_3) \psi_r(x_4) 
\rangle\\ 
\nonumber&=& 
(x_3-x_1)^{-h_m-h_n+h_r}(x_4-x_3)^{h_m-h_n-h_r}(x_4-x_1)^{-h_m+h_n-h_r}F\left(\frac{(x_2-x_1)(x_4-x_3)}{(x
_3-x_1)(x_4-x_2)} \right) \;,
\eea
and pick $x_1 < x_2 < x_3 < x_4$,  for later convenience. The $\psi_1$ null-vector 
\be
\left(3L­_{-1}^2-2L_{-2}\right)\psi_1
\ee
implies that this function obeys the differential equation
\be
0=\left( 2 \sum_{i \in \{1,3,4\}}\left( \frac{h_i}{(x­_i-x­_2)^2}-\frac{1}{x­_i-x­_2}\;\frac{\p}{\p x_i}\right)-3\left(\frac{\p}{\p x_2} \right)^2 \right)\mathcal{G}_{m 1,n r}(x_1,x_2,x_3,x_4)\; ,
\ee
where $h_1=h­_m, h_3=h­_n$, and $h_4=h­_r$.

Exploiting conformal symmetry by letting $\{x_1,x_2,x_3,x_4\} \to \{0,\l,1,\infty\}$ leads to
\be\label{GenDE}
0=F''(\l)+\frac{2(1-2 \l)}{3 \l (1-\l)} F'(\l)+\frac{2(\l(1-\l)h_r-\l h_n-(1-\l)h_m)}{3 \l^2 
(1-\l)^2}F(\l)\; .
\ee
Note that our conditions on the $x_i$ imply $0 < \l < 1$.

Equation (\ref{GenDE}) is linear, homogeneous, and has three regular singular points at $0$, $1$ and 
$\infty$.  Thus it can be transformed into a Riemann-Papperitz form with hypergeometric solutions 
\cite{Temme}.  We let 
\be \label{RPform}
F(\l)=\l^{m/3}(1-\l)^{n/3}f(\l)
\ee 
and insert the explicit forms of the weights (at $c=0$), giving
\be\label{TransDE1}
0=f''(\l)+\frac{2(1-2\l+m(1-\l)-n \l)}{3 \l (1-\l)} f'(\l)-\frac{(m+n)(m+n+1)+r(1-r)}{9 \l (1-\l)}f(\l)\; 
. 
\ee
(Note that up to (\ref{GenDE}) the analysis applies to the minimal models generally; only in (\ref{TransDE1}) do we 
specialize to $c=0$.)
Defining
\be
a=\frac{m+n+r}{3},\quad b=\frac{m+n+1-r}{3},\quad \textrm{and} \quad c=\frac{2m+2}{3}\; ,
\ee
we recover the hypergeometric differential equation
\be \label{hge}
0=\l(1-\l)f''(\l)+\left( c-(1+a+b)\l \right) f'(\l)-a b f(\l)\; .
\ee

Using the two solutions of (\ref{hge}) and (\ref{RPform})  gives the two conformal blocks  
\bea \label{mp1lk}
\mathcal{F}_{m 1, n r}^{m+1}(\lambda) &=& \lambda^{m/3} (1-\lambda)^{n/3} {}_2F_1\left(\begin{array}{c|}\frac{m+n+r}{3},\quad \frac{m+n+1-r}{3}  \\ \frac{2(m+1)}{3}\end{array}\quad\lambda \right)  \;\;\textrm{and}\\
%\nonumber &=& \lambda^{m/3} (1-\lambda)^{(1-n)/3} {}_2F_1\left( \begin{array}{c|}\frac{2+m-n-r}{3},\quad \frac{m+1-n+r}{3}\\ \frac{2(m+1)}{3}\end{array}\quad \lambda \right)\; ,\\ 
\label{mm1lk} \mathcal{F}_{m 1, n r}^{m-1}(\lambda) &=& \lambda^{(1-m)/3} (1-\lambda)^{n/3} {}_2F_1\left(\begin{array}{c|}\frac{1-m+n+r}{3},\quad \frac{2-m+n-r}{3}\\ \frac{2(2-m)}{3}\end{array}\quad \lambda \right)\; .
%\nonumber &=& \lambda^{(1-m)/3} (1-\lambda)^{(1-n)/3} {}_2F_1\left(\begin{array}{c|}\frac{3-m-n-r}{3},\quad \frac{2-m-n+r}{3}\\ \frac{2(2-m)}{3}\end{array}\quad \lambda \right)\; .
\eea
When   $m \equiv 2 \pmod{3}$   the second hypergeometric solution (Eq. (\ref{mm1lk})) must be modified 
because of logarithmic effects, reflecting the logarithmic nature of CFT for $c=0$. (Mathematically, this 
occurs because the parameter $\frac{2(2-m)}{3} \in -2{\mathbf N}$.)  However, this does not affect the 
validity of (\ref{mp1lk}) since for blocks with weights differing by integers, the block with the 
highest-order leading term is regular  \cite{G93}.  (Four-point functions involving $\psi_1$ are also 
analyzed in \cite{DF}, but this work does not consider the logarithmic case. )

Thus (\ref{mp1lk})  and  (\ref{mm1lk})  give the (non-logarithmic) conformal blocks for an arbitrary 
four-point correlation function (with one $\psi_1$) of the $c=0$ first column boundary operators $\psi_s$. 
 (Explicit expressions for the null-vectors of these operators are given in \cite{BYB}.)

The crossing symmetry of conformal block (\ref{mp1lk}) follows on using standard hypergeometric identities 
found, for example, in \cite{AS}.   If we let $m \leftrightarrow n$ and $\l \to 1-\l$ in (\ref{mp1lk}) and 
(\ref{mm1lk}) the crossing symmetry is expressed in terms of the $\l \to 1$ blocks as 
\be\label{XSymGood}\mathcal{F}_{m 1, n r}^{m+1}(\lambda) = \frac{\gf{2 m+2}\gf{2n-1}}{\gf{m+n+r}\gf{m+n+1-r}}\mathcal{F}_{n 1, m r}^{n-1}(1-\lambda)+\frac{\gf{2m+2}\gf{1-2n}}{\gf{m+2-n-r}\gf{m+1-n+r}}\mathcal{F}_{n 1, m r}^{n+1}(1-\lambda) \;.
\ee
This allows comparison with (\ref{GenXSym}), which in this case becomes
\be
C_{1, m, m+1} C_{r, n, m+1}\mathcal{F}_{m 1, n r}^{m+1}(\l)=C_{1, n-1, n} C_{r, n-1, m}\mathcal{F}_{n 1, m 
r}^{n-1}(1-\l)-C_{1, n, n+1} C_{r, n+1, m}\mathcal{F}_{n 1, m r}^{n+1}(1-\l)\; .
\ee

Before proceeding further, note that for $n \equiv 2 \pmod{3}$ there are logarithmic terms in the 
expansion of (\ref{mp1lk}) around $\l=1$.  In this case the crossing symmetry relation is
\bea \nonumber
\mathcal{F}_{m 1, n r}^{m+1}(\lambda) &=& \frac{\gf{2 m+2}\gf{2n-1}}{\gf{m+n+r}\gf{m+n+1-r}}\l^{m/3}(1-\l)^{(1-n)/3} \sum_{q=0}^{\frac{2n-4}{3}}\frac{\ph{m+1-n+r}\ph{m+2-n-r}}{q!\ph{4-2n}}(1-\l)^q\\
\label{XSymLog}&&\quad +\frac{\gf{2 m+2}}{\gf{m+1-n+r}\gf{m+2-n-r}}\l^{m/3}(1-\l)^{n/3}\sum_{q=0}^{\infty}\frac{\ph{m+n+r}\ph{m+n+1-r}}{q!\left( q +\frac{2n-1}{3} \right)!}(1-\l)^q\\
\nonumber &&\qquad \times \left( \log \left( 1-\l \right)-\psi \left( q+1 \right)-\psi \left( q+\textstyle 
\frac{2n+2}{3}\displaystyle \right)+\psi \left( q+\textstyle \frac{m+n+r}{3} \displaystyle \right)+\psi 
\left( q+\textstyle \frac{m+n+1-r}{3} \displaystyle \right) \right)\; ,
\eea
with $\psi$ the digamma function, instead of (\ref{XSymGood}).

 By comparison with (\ref{genOPE}) we see that the exponent of the leading term in (\ref{XSymLog}) belongs 
to the $n-1$ block.  Since the coefficient of this block is the same in (\ref{XSymGood}) as in 
(\ref{XSymLog}), we find that, regardless of logarithmic terms, the ratio of coefficients is
\be \label{main1}
\mathcal{R}_{m, n, r}=\frac{C_{1,n-1,n}C_{r,n-1,m}}{C_{1,m,m+1}C_{r,n,m+1}}=\frac{\gf{2 
m+2}\gf{2n-1}}{\gf{m+n+r}\gf{m+n+1-r}}\; .
\ee
There are four possible coefficient ratios that can be deduced from crossing symmetry, but (\ref{main1}) 
alone is sufficient to determine the OPE coefficients, so we omit the others.  

Now (\ref{main1}) gives
\be\label{main2}
\frac{1}{\sqrt{\mathcal{R}_{m, n, r}\mathcal{R}_{n-1, m+1, r}}}=\frac{C_{r,n,m+1}}{C_{r,n-1,m}} =\frac{ 
\gf{m+n+r}\gf{m+n+1-r} }{\sqrt{\gf{2m+1}\gf{2 m+2}\gf{2n-1}\gf{2 n}}}\; ,
\ee
a form which we will use to determine the OPE coefficients.  

We construct a general expression for $C_{m n r}$, with $0 < m \leq n \leq r$, by 
using (\ref{main2}) to increase the indices in pairs. (Since we are free to permute the indices of $C_{m n 
r}$, and the coefficients with an index $0$ are given by
\be \label{unit}
C_{0 m m} = 1 \; ,
\ee
this implies no loss of 
generality.)  It follows from (\ref{fus}) that $m+n+r$ is even,  and thus $(m+n-r)/2  \in {\mathbf N}$.  
Now
\be
C_{\frac{m+n-r}{2},\frac{m+n-r}{2},\,0}=1 \;.
\ee
Next, apply (\ref{main2}) to raise the second and third indices $(m-n+r)/2$ times. This results in
\be
C_{\frac{m+n-r}{2}, m, 
\frac{m-n+r}{2}}=\prod_{k=1}^{\frac{m-n+r}{2}}\sqrt{\frac{\gf{m+n-r+2k-1}\gf{2k}}{\gf{m+n-r+2k}\gf{2k-1}}}
\; .
\ee
Exchanging the first and second indices, and then applying (\ref{main2}) an additional $(-m+n+r)/2$ times, 
gives 
\bea \nonumber
C_{m n r} &=& \left(\prod_{\ell=1}^{\frac{-m+n+r}{2}}\frac{\gf{2m+2\ell -1}\gf{2 \ell}}{\sqrt{\gf{m-n+r+2 
\ell}\gf{m+n-r+2 \ell}\gf{m-n+r+2 \ell-1}\gf{m+n-r+2 \ell-1}}}\right)\\ \label{GenOPEForm}
&&\qquad \qquad \times 
\left(\prod_{k=1}^{\frac{m-n+r}{2}}\sqrt{\frac{\gf{m+n-r+2k-1}\gf{2k}}{\gf{m+n-r+2k}\gf{2k-1}}}\right)\; .
\eea

Note that the order in which we used (\ref{main2}) to raise pairs of indices and obtain (\ref{GenOPEForm}) 
is not unique; other orderings give different but equivalent expressions.  

Further manipulations of (\ref{GenOPEForm}) lead to an expression that reflects the symmetry of the $C_{m 
n r}$ but is somewhat more complicated since it involves more products,
\bea 
C­_{m, n, r}=\frac{
\left( \prod_{j=1}^{\frac{m+n+r}{2}}\gf{2 j -1} \right) 
\left( \prod_{j=1}^{\frac{m+n-r}{2}}\gf{2 j} \right)
\left( \prod_{j=1}^{\frac{m-n+r}{2}}\gf{2 j} \right)
\left( \prod_{j=1}^{\frac{-m+n+r}{2}}\gf{2 j} \right)}
{\sqrt{
\left( \prod_{j=1}^{2m}\gf{j} \right)
\left( \prod_{j=1}^{2n}\gf{j} \right)
\left( \prod_{j=1}^{2r}\gf{j} \right)}} \;. \label{NewGenOPEForm}
\eea

Equations (\ref{GenOPEForm}) and (\ref{NewGenOPEForm}) are the main results of this section.  They provide  closed expressions for all OPE 
coefficients between operators in the first column of the Kac table at $c=0$. 

Considering particular indices can lead to more compact expressions.  In the following, we find the 
lowest-order cases, i.e.  $C_{mnr}$ with $m=1, 2, 3$, directly from  (\ref{main2}). 

Setting $n = 1$ and $r = m$ in (\ref{main2}) gives 
\be\label{1mmp1Res}
C_{1,m,m+1}=\sqrt{\frac{\gf{2 m+1}\gf{2}}{\gf{2 m+2}\gf{1}}} \;.
\ee
Setting $m \to m-1$ in (\ref{1mmp1Res}) and permuting indices gives $C_{1,m,m-1}$, which (as follows from  
(\ref{fus})) exhausts the possibilities for $C_{1,m,r}$. If we let $n = 2$ and $r = m+1$ in (\ref{main2}), 
include result (\ref{1mmp1Res}), and redefine $m \to m-1$ we find 
\be\label{2mmRes}
C_{2 ,m ,m} = \frac{\gf{2 m+1}\gf{2}}{\gf{2 m}}\sqrt{\frac{\gf{2}}{\gf{4}\gf{1}}}\; .
\ee
If we let $n = 2$ and $r = m-1$ in (\ref{main2}), redefine $m \to m+1$, and include result 
(\ref{1mmp1Res}) we find 
\be\label{2mmp2Res}
C_{2, m, m+2} = \sqrt{\frac{\gf{2 m+3}\gf{2}}{(2 m+1)\gf{2 m+2}}}\; .
\ee
Note that, by letting $m \to m-2$, (\ref{2mmp2Res}) also includes the coefficient $C_{2, m-2, m}$, which 
is the only other term in the fusion $\psi_m \psi_2$.  The expressions in (\ref{3mmp1Res}) and 
(\ref{nexteq}) below can be similarly manipulated to give all the non-vanishing $C_{3mr}$.

If we let $n = 3$ and $r = m$ in (\ref{main2}) and include result (\ref{2mmRes}) we find 
\be\label{3mmp1Res}
C_{3, m, m+1} = \frac{m \gf{2}}{3} \sqrt{\frac{2 \gf{2 m+1}}{\gf{2m+2}}}\; .
\ee
If we let $n = 3$ and $r = m-2$ in (\ref{main2}), redefine $m \to m+2$, and include result 
(\ref{2mmp2Res}) we find 
\be\label{nexteq}
C_{3, m, m+3} = \sqrt{\frac{3(m+1)}{(2m+1)(2m+3)}}\; .
\ee
Interestingly, this is a purely algebraic number.  

The results (\ref{1mmp1Res}-\ref{nexteq}) also follow, as they must, from (\ref{GenOPEForm}) or 
(\ref{NewGenOPEForm}). The procedure leading to (\ref{1mmp1Res}-\ref{nexteq}) can be 
generalized to obtain other coefficients in simpler forms  than   (\ref{GenOPEForm}) or (\ref{NewGenOPEForm}). 

The three coefficients used explicitly in \cite{KSZ,SKZ,SKZip} are
\be
C_{1 1 2} = \sqrt{\frac{\gf{2}}{\gf{4}\gf{1}}}  = \frac{2^{1/2}\; \pi^{1/2} \;3^{1/4}}{\G(1/3)^{3/2}}\; ,
\ee
which follows immediately from  (\ref{1mmp1Res}) or  (\ref{2mmRes}) (and $\G$ function identities),
\be
C_{1 2 3} = \sqrt{\frac{\gf{5}\gf{2}}{\gf{1}}} = \frac{2^{3/2}\; \pi}{3\; \G(1/3)^{3/2}} \; ,
\ee
from (\ref{1mmp1Res}), and
\be\label{C222}
C_{2 2 2} = \frac{\gf{5}\gf{2}}{\gf{4}}\sqrt{\frac{\gf{2}}{\gf{4}\gf{1}}} = \frac{2^{7/2}\; 
\pi^{5/2}}{3^{3/4}\; \G(1/3)^{9/2}} \; ,
\ee
from (\ref{2mmRes}).

The square of $C_{1 1 2}$ appears as the leading coefficient in Cardy's formula \cite{JC1}; our form 
agrees with Cardy's result.  The ratio $C_{1 1 2}/ C_{1 2 3}$ is a coefficient in a formula in \cite{SKZ}.
The  coefficient $C_{2 2 2}$ appears as a proportionality constant in \cite{KSZ}.  All of these 
coefficients are excellent agreement with simulations of the formulas in which they appear, for example  
$C_{2 2 2} = 1.02993\ldots$ as compared to $1.030 \pm 0.001$ from numerical results \cite{KSZ}.

%SECTION
\section{Generalization to Minimal Models} \label{min}

In this section, by means of a straightforward generalization of section \ref{c=0}, we calculate conformal 
blocks and operator product expansion coefficients for the (augmented) minimal models, i.e. we include coefficients for
non-unitary minimal models.  In the unitary case, many of the results that we derive are 
already known \cite{Lew}.  However, our extension to the augmented minimal models is to our knowledge new.  As a 
consequence, some of our results are useful in calculating  correlation functions of Fortuin-Kasteleyn clusters in $q$-state Potts models \cite{SSKZ}.
(For spin clusters in these cases  the corresponding OPE coefficients may vanish.)  In what follows, we present closed form expressions for the coefficients that are valid when $\phi_{1,2}$ mediates a change from fixed to free boundary conditions. (Actually, our results do not specifically require fixed or free boundary conditions.  They also apply whenever only two boundary conditions, of any type, enter the problem and $\phi_{1,2}$ specifies the points where they abut.)

To avoid confusion with the index $m$ used above, we denote the minimal model parameter by $M$, so that 
the central charge is
\be
c=1-\frac{6}{M(M+1)}\; .
\ee
The boundary operators considered, as above, are generated by products of $\psi_1 := \phi_{1,2}$, namely  
\be
\psi_n := \phi_{1, n+1}, \quad \mathrm{with\; weights} \quad h_n = \frac{n(n M-2)}{4(M+1)}\; .
\ee
Note that for the unitary  minimal models the index n is restricted to $0 \leq n < M$. In this case $M$ is an integer, but in fact our results apply for any real $M$ with $M \ge 2$.

We start with the four-point function, now including the boundary condition indices
\be
\mathcal{G}_{m 1, n r}^{a b c, c d a}(x_1,x_2,x_3,x_4):=\langle \psi_m^{a b}(x_1) \psi_1^{b c}(x­_2) 
\psi_n^{c d}(x_3) \psi_r^{d a}(x­_4) \rangle\; .
\ee

Proceeding as in section \ref{c=0}, from (\ref{cfform}) through  (\ref{hge}), we find the conformal blocks 
(the boundary condition dependence of the correlation function enters only via the OPE coefficients, the 
blocks themselves being independent of boundary conditions)
\bea \label{genmp1lk}
\mathcal{F}_{m 1, n r}^{m+1}(\lambda) &=& 
\lambda^{\frac{m M}{2(M+1)}} (1-\lambda)^{\frac{n M}{2(M+1)}} 
{}_2F_1\left( \begin{array}{c|}\frac{M(1+m+n+r)-2}{2(M+1)},\quad \frac{2(1+m+n-r)}{2(M+1)}\\ 
\frac{M(m+1)}{M+1}\end{array}\quad \lambda \right)\\
%\nonumber &=& 
%\lambda^{\frac{m M}{2(M+1)}} (1-\lambda)^{\frac{2-n M}{2(M+1)}} 
%{}_2F_1\left( \begin{array}{c|}\frac{M(1+m-n-r)+2}{2(M+1)},\quad \frac{M(1+m-n+r)}{2(M+1)}\\ 
%\frac{M(m+1)}{M+1}\end{array}\quad \lambda \right)\; ,\\
\mathcal{F}_{m 1, n r}^{m-1}(\lambda) &=& \label{genmm1lk}
\lambda^{\frac{2-m M}{2(M+1)}} (1-\lambda)^{\frac{n M}{2(M+1)}} 
{}_2F_1\left(\begin{array}{c|}\frac{M(1-m+n+r)}{2(M+1)},\quad \frac{M(1-m+n-r)+2}{2(M+1)}\\ 
\frac{2+M(1-m)}{M+1}\end{array}\quad \lambda \right)
%\\ \nonumber &=&
%\lambda^{\frac{2-m M}{2(M+1)}} (1-\lambda)^{\frac{2-n M}{2(M+1)}} 
%{}_2F_1\left(\begin{array}{c|}\frac{4+M(1-m-n-r)}{2(M+1)},\quad \frac{2+M(1-m-n+r)}{2(M+1)}\\ 
%\frac{2+M(1-m)}{M+1}\end{array}\quad \lambda \right)\; .
\eea
The Kac fusion rules impose the conditions
\bea
m+n+r &\equiv& 1 \pmod{2}\\
n+r \;\; \geq \;\; m \pm 1  &\geq& |n-r|\; .
\eea
For certain values of $m$ and $M$, $\mathcal{F}_{m 1, n r}^{m-1}(\lambda)$ is a logarithmic conformal block.  This situation arises when $(2+M(1-m))/(M+1)$ is a non-positive integer,  so that (\ref{genmm1lk}) is undefined.  However  (\ref{genmp1lk}), which is the equation that we use, remains valid.

In the non-logarithmic case the  crossing symmetry relations give
\bea
\nonumber \mathcal{F}_{m 1, n r}^{m-1}(\lambda) &=& 
\frac{\gfM{2-M(m-1)}\gfM{M n-1}}{\gft{M(1-m+n+r)}\gft{M(1-m+n-r)+2}}\mathcal{F}_{n 1, m 
r}^{n-1}(1-\lambda)\\
&&+\frac{\gfM{2-M(m-1)}\gfM{1-M n}}{\gft{4+M(1-m-n-r)}\gft{2+M(1-m-n+r)}}\mathcal{F}_{n 1, m 
r}^{n+1}(1-\lambda)\\
\nonumber \mathcal{F}_{m 1, n r}^{m+1}(\lambda) &=& 
\frac{\gfM{M(m+1)}\gfM{M n-1}}{\gft{M(1+m+n+r)-2}\gft{M(1+m+n-r)}}\mathcal{F}_{n 1, m r}^{n-1}(1-\lambda)\\
&&+\frac{\gfM{M(m+1)}\gfM{1-M n}}{\gft{2+M(1+m-n-r)}\gft{M(1+m-n+r)}}\mathcal{F}_{n 1, m r}^{n+1}(1-\lambda)
\eea

As before we compare these crossing relations to the generic form
\bea \label{bclabXrel}
C_{m, 1, m+1}^{ a b c} C_{n, r, m+1}^{c d a}\alpha_{m+1}^{a c}\mathcal{F}_{m 1, n r}^{m+1}(\l)&=&C_{n, 1, 
n-1}^{d c b} C_{m,r, n-1}^{b a d}\alpha_{n-1}^{b d}\mathcal{F}_{n 1, m r}^{n-1}(1-\l)\\ \nonumber
&&-C_{n, 1, n+1}^{d c b} C_{m, r, n+1}^{b a d} \alpha_{m+1}^{b d}\mathcal{F}_{n 1, m r}^{n+1}(1-\l)\; .
\eea
As explained above,  except for percolation ($M=2$) we must include the boundary condition labels on the OPE coefficients.  We also need (as explained below) the $\alpha$ factors, which encode the normalization of the boundary two point function via 
\be
\langle \psi_j^{a b}(x_2) \psi_j^{b a}(x_1) \rangle = \alpha_j^{a b}(x_2-x_1)^{-2 h_j}\; .
\ee
Thus the three point function is
\be
\langle \psi_i^{a b}(x_3) \psi_j^{b c}(x_2) \psi_k^{c a}(x_1) \rangle = C_{i j k}^{a b c} \alpha_k^{a 
c}(x_2-x_1)^{h_i-h_j-h_k}(x_3-x_1)^{h_j-h_i-h_k}(x_3-x_2)^{h_k-h_i-h_j}\; .
\ee
Symmetry under the exchange of operators in the three point function leads to the relations
\be \label{cosyms}
C_{i j k}^{a b c} \alpha_k^{a c}=
C_{j i k}^{c b a} \alpha_k^{a c}=
C_{k i j}^{c a b} \alpha_j^{b c}=
C_{i k j}^{b a c} \alpha_j^{b c}=
C_{j k i}^{b c a} \alpha_i^{a b}=
C_{k j i}^{a c b} \alpha_i^{a b}\; .
\ee
When focusing on boundary operators, we could absorb the $\alpha$ coefficients into the normalization of the boundary operators as in section \ref{c=0}.  However the usual convention is to normalize the bulk operators  to unity.  Adopting that here keeps our results consistent with the literature \cite{Lew}, \emph{e.g.} the unitary Ising model boundary OPE coefficients.  Methods for evaluating the $\alpha$s are given in \cite{Lew}.

Thus we find
\be \label{Yeah}
\frac{\gfM{M(m+1)}\gfM{M n-1}}{\gft{M(1+m+n+r)-2}\gft{M(1+m+n-r)}}=\frac{C_{n, 1, n-1}^{d c b}C_{m, r, 
n-1}^{b a d} \alpha_{n-1}^{b d}}{C_{m, 1, m+1}^{a b c}C_{n, r, m+1}^{c d a} \alpha_{m+1}^{a c}}\; .
\ee
We focus on this particular relation because  it is the only one of the four possible coefficient ratios that does not need to be modified in the logarithmic case. This can be shown by a series expansion analogous to that employed in section \ref{anal}.  Thus (\ref{Yeah}) applies to all allowed $m, n$ and $r$ values.

Using (\ref{cosyms}) and (\ref{Yeah}) we can write down the generalization of (\ref{main2}) for 
$M \neq 2$
\be \label{Fin}
\frac{K_{m+1, n+1, r}^{a d c}}{K_{m, n, r}^{a b c}}  =
\frac{C_{m+1, n+1, r}^{a d c}}{C_{m, n, r}^{a b c}} \sqrt{\frac{\alpha_{m}^{a b}\alpha_{n}^{b 
c}}{\alpha_{m+1}^{a d}\alpha_{n+1}^{c d}}}  = 
\frac{\gft{M(2+m+n+r)-2}\gft{M(2+m+n-r)}}{\sqrt{\gfM{M(m+1)}\gfM{M(m+1)-1}\gfM{M(n+1)}\gfM{M(n+1)-1}}}\; .
\ee
To simplify notation we have defined
\be
K_{m, n, r}^{a b c} := C_{m, n, r}^{a b c}\sqrt{\frac{\alpha_{r}^{a c}}{\alpha_{m}^{a b}\alpha_{n}^{b c}}}\; ,
\ee
which is symmetric under simultaneous permutation of boundary conditions and operator labels,
\be
K_{m, n, r}^{a b c} = K_{n, r, m}^{b c a} = K_{r, m, n}^{c a b} = K_{n, m, r}^{c b a} = K_{r, n, m}^{a c b} = K_{m, r, n}^{b a c}\; .
\ee
We can use (\ref{Fin}) to raise pairs of operator indices on $K_{m, n, r}^{a b c}$.  

From this point on, we consider only cases in which $\psi_1$ mediates changes between free and fixed boundary conditions; thus all our boundary conditions are free or fixed. As a consequence, raising indices with (\ref{Fin}) changes the type of intermediate boundary condition from one to the other.

Using (\ref{Fin}) on $K_{m, n, r}^{a b c}$ we can derive the relations
\be
\frac{K_{m n r}^{a b c}}{K_{\frac{m-n+r}{2} \frac{-m+n+r}{2} r}^{a d c}} = \left(\prod_{\ell=1}^{\frac{m+n-r}{2}}\frac{\gfM{M(r+\ell) -1}\gfM{M \ell}}{\sqrt{\gft{M(m-n+r+2\ell)}\gft{M(m-n+r+2\ell)-2}\gft{M(-m+n+r+2\ell)}\gft{M(-m+n+r+2 \ell)-2}}}\right)\; ,
\ee
and
\be
\frac{K_{\frac{m-n+r}{2} \frac{-m+n+r}{2} r}^{a d c}}{K_{0 \frac{-m+n+r}{2} \frac{-m+n+r}{2}}^{d d c}} = \left(\prod_{k=1}^{\frac{m-n+r}{2}}\sqrt{\frac{\gft{M(-m+n+r+2k)-2}\gfM{M k}}{\gft{M(-m+n+r+2k)}\gfM{M k-1}}}\right)\; .
\ee

The fact that $C_{0 j j}^{a a b}=1$, for all $\psi_j$, $a$ and $b$ then implies that
\be\label{objterm}
K_{0 \frac{-m+n+r}{2} \frac{-m+n+r}{2}}^{d d c}=C_{0 \frac{-m+n+r}{2} \frac{-m+n+r}{2}}^{d d c}\sqrt{\frac{\alpha_{\frac{-m+n+r}{2}}^{d c}}{\alpha_{0}^{d d}\alpha_{\frac{-m+n+r}{2}}^{d c}}}=\frac{1}{\sqrt{\alpha_{0}^{d d}}}\; ,
\ee
and allows us to generalize (\ref{GenOPEForm}) and (\ref{NewGenOPEForm}).

Thus we find
\bea \nonumber
C_{m n r}^{a b c} &=&
\sqrt{\frac{\alpha_m^{a b}\alpha_n^{b c}}{\alpha_r^{a c}\alpha_0^{d d}}} 
\left(\prod_{\ell=1}^{\frac{m+n-r}{2}}\frac{\gfM{M(r+\ell) -1}\gfM{M \ell}}{\sqrt{\gft{M(m-n+r+2\ell)}\gft{M(m-n+r+2\ell)-2}\gft{M(-m+n+r+2\ell)}\gft{M(-m+n+r+2 \ell)-2}}}\right)\\ \label{GenGenOPEForm}
&&\qquad \qquad \times 
\left(\prod_{k=1}^{\frac{m-n+r}{2}}\sqrt{\frac{\gft{M(-m+n+r+2k)-2}\gfM{M k}}{\gft{M(-m+n+r+2k)}\gfM{M k-1}}}\right)\; ,
\eea
but (\ref{GenGenOPEForm}) can once again be rearranged as
\bea 
C­_{m, n, r}^{a b c}=
\sqrt{\frac{\alpha_m^{a b}\alpha_n^{b c}}{\alpha_r^{a c}\alpha_0^{d d}}}
\frac{
\left( \prod_{j=1}^{\frac{m+n+r}{2}}\gfM{M j -1} \right) 
\left( \prod_{j=1}^{\frac{m+n-r}{2}}\gfM{M j} \right)
\left( \prod_{j=1}^{\frac{m-n+r}{2}}\gfM{M j} \right)
\left( \prod_{j=1}^{\frac{-m+n+r}{2}}\gfM{M j} \right)}
{\sqrt{
\left( \prod_{j=1}^{m}\gfM{Mj-1}\gfM{Mj} \right)
\left( \prod_{j=1}^{n}\gfM{Mj-1}\gfM{Mj} \right)
\left( \prod_{j=1}^{r}\gfM{Mj-1}\gfM{Mj} \right)}}\; . \label{NewGenGenOPEForm}
\eea
The question of whether the boundary condition $d$ is free or fixed remains.  By comparison with (\ref{objterm}) we can see that if $(-m+n+r)/2$ is even (odd) than $d$ represents the same (opposite) boundary state as $c$.  For spin models, one can show that $d$ is fixed (free) if there is (not) a Fortuin-Kasteleyn cluster that connects all three points.

We list a few simple OPE coefficients that follow from (\ref{NewGenGenOPEForm}),
\bea
C_{1 1 2}^{f F f}\frac{\sqrt{\alpha_0^{F F}\alpha_2^{f f}}}{\alpha_1^{F f}}\: =\: C_{1 1 2}^{F f F}\frac{\sqrt{\alpha_0^{f f}\alpha_2^{F F}}}{\alpha_1^{F f}}&=&\sqrt{\frac{\gfM{2M-1}\gfM{M}}{\gfM{2 M}\gfM{M-1}}}\; ,\\
C_{1 2 3}^{f F F}\sqrt{\frac{\alpha_0^{F F}\alpha_3^{f F}}{\alpha_2^{F F}\alpha_1^{F f}}}\: =\:C_{1 2 3}^{F f f}\sqrt{\frac{\alpha_0^{f f}\alpha_3^{f F}}{\alpha_2^{f f}\alpha_1^{f F}}}&=& \sqrt{\frac{\gfM{3M-1}\gfM{M}}{\gfM{3 M}\gfM{M-1}}}\; ,\\
C_{2 2 2}^{f f f}\sqrt{\frac{\alpha_0^{F F}}{\alpha_2^{f f}}}\: =\: C_{2 2 2}^{F F F}\sqrt{\frac{\alpha_0^{f f}}{\alpha_2^{F F}}} &=&\sqrt{\frac{\gfM{3M-1}^2\gfM{M}^3}{\gfM{2 
M}^3\gfM{M-1}\gfM{2 M-1}}}\; .
\eea
Here $f$ ($F$) denotes free (fixed) boundary conditions.  We do not need to differentiate between boundaries fixed in {\it different} spin states, since by symmetry the OPE coefficients must be independent of this.  See \cite{Lew} for more information on the $\alpha$s. One finds, for example, using the results of \cite{Lew} for the necessary $\alpha$ values, for Ising clusters ($M=3$), that $C_{2 2 2}^{f f f}=1.52552\dots$ and $C_{2 2 2}^{FFF}=1.07871\dots$.

%SECTION
\section{Conclusions}

This paper derives  closed form expressions for arbitrary Kac table boundary operator product expansion 
coefficients $C_{mnr}$ involving first-column operators $\phi_{1,s}$ in the $c=0$  conformal field theory.  Similar results for the coefficients $C^{abc}_{mnr}$ when $\phi_{1,2}$ mediates a change from fixed to free (or any two specified)  boundary conditions
in both unitary and augmented minimal models are also given.  These coefficients occur in a variety of formulas for two-dimensional 
critical systems,  derived via conformal field theory. These coefficients may also play a role in further 
exploring  $c=0$ conformal field theory through descendant operator relations such as those in \cite{SKZ}.

%SECTION
\section{Acknowledgments}\label{ack}
We thank W. O. Bray, J. Cardy and R. Maier for useful suggestions. 

This work was supported in part by the National Science Foundation Grant No. DMR-0203589 and continuation 
grant DMR-0536927 (PK) and by EPSRC Grant No. EP/D070643/1 (JJHS).

PK is grateful for the hospitality of the Rudolf Peierls Centre for Theoretical Physics, Oxford University, where part of this work was performed.

\end{document}